# Estimating Mixed-Mode Urban Trail Traffic Using Negative Binomial Regression Models

Xize Wang[a, *], Greg Lindsey[b], Steve Hankey[c], Kris Hoff[d]

## Abstract

Data and models of non-motorized traffic on multiuse urban trails are needed to improve planning and management of urban transportation systems. Negative binomial regression models are appropriate and useful when dependent variables are non-negative integers with over-dispersion like traffic counts. This paper presents eight negative binomial models for estimating urban trail traffic using 1,898 daily mixed-mode traffic counts from active infrared monitors at six locations in Minneapolis, MN. Our models include up to 10 independent variables that represent socio-demographic, built environment, weather, and temporal characteristics. A general model can be used to estimate traffic at locations where traffic has not been monitored. A six-location model with dummy variables for each monitoring site rather than neighborhood specific variables can be used to estimate traffic at existing locations when counts from monitors are not available. Six trail-specific models are appropriate for estimating variation in traffic in response to variations in weather and day of week. Validation results indicate negative binomial models outperform models estimated by ordinary least squares regression. These new models estimate traffic within approximately 16.3% error, on average, which is reasonable for planning and management purposes.

a. Price School of Public Policy, University of Southern California. Email: wangxize316@gmail.com. OCRID: 0000-0002-4861-6002.
b. Humphrey School of Public Affairs, University of Minnesota. Email: linds301@umn.edu.
c. College of Science and Engineering, University of Minnesota. Email: shankey1028@gmail.com. OCRID: 0000-0002-7530-6077.
d. National Community Stabilization Trust, Bloomington, MN.
*. Corresponding author.

**INTRODUCTION**

Non-motorized transportation includes travel by bicycle or walking. Although engineers and planners have worked to measure and model non-motorized transportation since the 1970s, traffic counts and other data required for planning and modeling remain inadequate and generally unavailable, especially compared to data and models for motorized traffic. With initiatives like Complete Streets (National Complete Street Coalition 2011) that promote the integration of facilities for all modes of transportation, the need for data and models for planning has increased.

Previous research has shown that the construction and maintenance of biking and walking facilities, including multiuse trails or shared-use paths, can increase work and non-work non-motorized travel (Bowman et al. 1994; Krizek et al. 2009). Increased non-motorized travel facilities offer a number of supplemental benefits to transportation systems and their users, including greater mode choice and reductions in travel time for short trips (Pedestrian and Bicycle Information Center 2010), environmental benefits such as reduced air pollution (Pedestrian and Bicycle Information Center 2010; Woodcock et al. 2009) and improved quality of life (Gobster 2005). In addition, non-motorized transportation is associated with increased physical activities (Librett et al. 2006). Infrastructure for active travel also yields economic benefits such as increased job creation (Garrett-Peltier 2011) and reduced medical expenditures (Wang et al. 2005). Because of these benefits, federal, state and local governments have increased funding in recent years for many programs to enhance non-motorized transportation systems. The Pedestrian and Bicycle Information Center (2010) estimates that funding for non-motorized transport projects increased from less than $50 million in 1992 to around $1.2 billion in 2009. Some of these increases in investment were in urban multi-use trails that are integrated into street networks.

An example of these new investments is the federally-sponsored Non-Motorized Transportation Pilot Program. In this program, the Federal Highway Administration (FHWA) has provided $25 million over five years to each of four pilot communities (including the Twin Cities metropolitan area) to support projects such as bicycle facility construction. Major goals of this program are to increase non-motorized traffic mode share and volumes and safety of cyclists and pedestrians (Pedestrian and Bicycle Information Center 2010). To assist with evaluation of the Pilot Program, the Minneapolis Department of Public Works and Transit for Livable Communities, a nonprofit organization that is responsible for implementation of the Pilot Program in the Twin Cities metropolitan area have initiated counts of non-motorized traffic. These counts include manual 2-hour field observations of both bicyclists and pedestrians at more than 250 locations on different types of non-motorized facilities and, of particular interest in this study, automated counts on a limited number of locations on multiuse trails (FHWA 2012; Hankey et al. 2012). Researchers at the University of Minnesota have augmented these counts with continuous active infrared counters that count mixed mode traffic – undifferentiated bicyclists and pedestrians – at six locations on multiuse trails. These counts have informed management decisions to increase traffic flow and safety. For example, the City of Minneapolis has changed signage at intersections of residential collector streets with a multi-use trail, the Midtown Greenway, to give trail users the right-of-way because trail traffic counts were greater than motorized volumes on the residential streets (Anderson 2010). More generally, counts and estimates of traffic on non-motorized facilities, including urban multiuse trails, are needed to inform investment decisions by local metropolitan planning organizations (MPOs).

The models in this paper, which estimate mixed-mode trail traffic, incorporate the categories of variables used in comparable models reported previously by different authors (e.g.,

Lindsey et al. 2007; Coutts 2008; Coutts 2009) but take a different functional form (i.e., the negative binominal) and are estimated from data sets from locations not previously reported. These models will be especially useful for informing planning and management decisions focusing on urban multiuse trails and can be used along with other models of bicycle and pedestrian traffic on streets and sidewalks, respectively, to provide more complete estimates of non-motorized traffic levels. Part 2 of this paper reviews recent studies that (a) explore correlates of urban bicycle and pedestrian traffic and (b) employ different methods for estimating and projecting existing and future bicycle traffic. Part 3 summarizes our methods for collecting trail traffic counts and our approach to analysis. Part 4 presents results, including a general model, a six-location model, and six trail-specific models for each monitoring location. Part 5 presents our validation of the models. Parts 6 and 7 review our main findings, note the limitations of this study, and suggest areas for future research.

**LITERATURE REVIEW**

Two main issues explored in the literature on urban non-motorized transportation are (a) increasing use and (b) improving safety. This review is limited to studies related to measuring and explaining variations in use or traffic volumes. Previous studies about traffic volume have focused on determining factors that impact bicycle and pedestrian traffic and incorporate a variety of different methods of data collection and analysis. This section reviews (a) variables shown to be correlated with measures of bicycle and pedestrian traffic and (b) methods used to estimate models.

**Correlates of Non-motorized Traffic**

Researchers have identified categories of variables that are correlated with urban non-motorized traffic: (a) socio-demographic characteristics, (b) built environment or urban form, including bicycle or pedestrian specific infrastructure, and (c) weather, and (d) temporal factors such as day of week. Many of these variables are correlated with both bicycles and pedestrian traffic but the strength of their association may vary. Researchers have modeled bicycle and pedestrian traffic both separately and combined as mixed-mode traffic depending on the type of counts available. For example, counts from inductive loop detectors enable modeling of bicycle traffic only while counts from infrared monitors permit only modeling of undifferentiated non-motorized traffic.

*Socio-demographic characteristics*

Bicycle and pedestrian traffic volumes, including volumes on trails, have been found to correlate with education, income, age, race and ethnicity, and student status (i.e., student vs. non-student). Some studies have focused on characteristics of users as reported in surveys while others have correlated traffic volumes with neighborhood characteristics where counts have been conducted. Moudon et al. (2005) found the likelihood of cycling increases for male and younger adults. Rodríguez and Joo (2004) found students tend to use bicycles more than non-students. Lindsey et al. (2007) reported that mixed mode trail traffic tended to be higher in neighborhoods with greater education, higher incomes, and with larger proportions of people over 64 years old and less than 5 years old. Hankey et al. (2012) modeled bicycle and pedestrian traffic separately from two hour field counts and found that traffic levels for both were associated with higher levels of education and higher proportions of non-white residents in nearby neighborhoods.

Bicycle traffic was negatively associated with neighborhood household income, while pedestrian traffic was not significantly associated with income.

*Built Environment, Urban Form, and Infrastructure*

The built environment may affect travel behavior in different ways (Ewing and Cervero 2010). Moudon et al. (2005) and Handy et al. (2002), for example, found higher population density is related to higher levels of bicycling. Other factors such as intersection density and land use mix have also been shown to increase rates of active travel (Frank et al. 2005). Forsyth et al. (2008) and Lindsey at al. (2007) found higher commercial land use percentage and lower street length are associated with increases in the number of pedestrian and mixed-mode traffic, respectively. Examples of urban bicycle infrastructure include bike lanes on streets, separate bike paths and "bicycle boulevards" (Dill 2009). In a cross-sectional analysis of 35 large US cities, Dill and Carr (2003) have shown that higher levels of bicycle infrastructure are associated with higher volumes of bicycle commuting travel. Schneider et al. (2009) showed that pedestrian intersection crossings were associated with land use characteristics and transportation infrastructure. Griswold et al. (2011) found that bicycle intersection volumes were associated with the number of nearby retail facilities and other land use characteristics.

*Weather*

Studies have demonstrated that non-motorized traffic varies systematically with daily and seasonal variations in weather, especially in temperate climates. Non-motorized traffic volumes have been shown to correlate inversely with precipitation and positively with temperature. The latter relationship is nonlinear, and traffic volumes are depressed with extremely high temperatures (Lindsey et al. 2007; Stinson and Bhat 2004) .In addition, traffic volumes vary in relation to expectations about weather, which are based on season or long-term averages. For example, Lindsey et al. (2007) have shown that, in temperate climates, a specific temperature (e.g., 10 degrees Celsius) can induce a spike in trail traffic in early spring but depress traffic in a late summer month when expectations based on average temperatures are that the temperature "should" be much higher.

*Temporal factors*

Several studies have demonstrated that non-motorized traffic varies systematically with temporal factors, including time of day, day of week, and month of year (Hunter and Huang 1995) and (Jones 2009). These variations depend in part on the purpose of traffic, which itself may be correlated with location and mode. That is, depending on its location in a transportation network, traffic may include different proportions of utilitarian or recreational trips, with corresponding differences in day of week traffic volumes (Nordback 2012). After controlling for variations in weather, location, and other factors, Lindsey and Nguyen (2004) and Lindsey et al. (2007) have shown that urban trail traffic often is higher on weekend days. This finding also has been reported by others.

**Methods of Analyzing Non-motorized Traffic**

Methods used to measure and explain variations in non-motorized traffic vary among studies. Researchers have used a variety of methods to collect data about levels of use and traffic volumes and different approaches to analyze and model traffic volumes.

*Data collection methods*

The two principal sources of data on non-motorized traffic are surveys of people and traffic volumes on infrastructure. The most commonly reported data are surveys about cycling behavior, including characteristics of cyclists. Surveys have been conducted in cities such as Minneapolis and St. Paul, MN (Krizek and Johnson 2006), Indianapolis, IN (Ottensmann and Lindsey 2008), Seattle, WA (Moudon et al. 2005) and Chapel Hill, NC (Rodríguez and Joo 2004), among others. A commonly used source of data on travel behavior, including mode of commuting, is from census surveys. Census surveys, which have been used in both the US and UK, can be combined with socio-demographic characteristics to understand how travel behavior correlates with these characteristics (e.g., Parkin et al., 2008). Travel diaries also have been used to collected detailed information on non-motorized transport.

Counts of traffic on non-motorized facilities can be obtained through a variety of methods, including in-field observations and deploying different types of automated counters such as magnetic loop detectors, active or passive infrared counters, video, or GPS devices. Dill (2009), for example, tracked the temporal-spatial biking activities of cyclists using GPS devices. Coutts (2009) collected similar data using a combination of field observation and GPS devices. Lindsey et al. (2007) used 30 TrailMaster© active infrared monitors to gather bicycle and pedestrian traffic data in Indianapolis, IN. Field counts through periodic visual observation remains the most common method of counting (Reynolds et al. 2007) and have been used to model bicycle and pedestrian traffic separately (e.g., Schneider et al. (2009); Griswold et al. (2011); Haynes and Andrzejewski (2010); Hankey et al. (2012)).

*Analytical methods*

Researchers also have used a variety of approaches and methods to analyze data on non-motorized transportation and estimate bicycle or pedestrian traffic (Porter et al., 1999). One approach, which we use here, is sometimes referred to as an aggregate-level method. Aggregate level methods involve analysis of trips and correlating them with the characteristics of an area through the use of quantitative techniques such as linear ordinary least squares (OLS) regression. While many studies have reported models estimated with OLS, researchers now are using nonlinear models such as negative binomial and Poisson regression techniques that have been used extensively in the modeling of traffic accidents (Miaou 1994). Cao et al. (2006)*,* for example, used negative binomial models to analyze how pedestrian behavior is affected by the built environment and by residential self-selection. Ottensmann and Lindsey (2008) used both logistic models and negative binomial models to predict trail use in Indianapolis, Indiana.

**DATA AND METHODS**

**Trail Traffic Counts**

Mixed-mode, non-motorized traffic was counted at six locations on multiuse trails in Minneapolis using TrailMaster© infrared monitors, which are reliable in counting trail traffic (Figure 1). Three locations are on the Midtown Greenway at intersections near arterial streets: Hennepin Avenue (#1), West River Parkway (#2) and Cedar Avenue (#3). The Midtown Greenway is a 5.5 mile multiuse trail developed on an historic rail line that is maintained by the City of Minneapolis Department of Public Works in collaboration with users and other stakeholders. The other three locations are on trails around lakes on property maintained by the Minneapolis Parks and Recreation Board: Lake Calhoun (#4), Lake Nokomis (#5) and Theodore

Wirth Park (#6). These six locations are in different neighborhoods with different built environment and socioeconomic characteristics.

[Figure 1 Here]

The active infrared counters record the time a user breaks an infrared beam between a transmitter and a receiver. Users may be either pedestrians or cyclists; the counters cannot distinguish between modes. Because users sometimes cross the beam simultaneously, the raw counts underestimate the actual trail traffic. To correct for this systematic error, the raw counts are aggregated by hour, and a correction equation developed by regressing TrailMaster© counts ($x$) on 130 hours of counts from field observations ($y$) is applied to each hourly total. The correction equation is:

$$y = 0.0002x^2 + 1.0655x - 1.2937 \quad (1)$$

The adjusted hourly totals then are aggregated to obtain a 24-hour daily total, which is the dependent variable in our models.

The total number of daily counts in our sample is 1,898; the sample includes counts taken from June 21, 2010 to September 23, 2011 (Table 1). The number of daily counts varies among the six locations. Hennepin Ave. has the most daily counts (427), while Nokomis Parkway has the fewest (261).

The mean daily traffic volumes vary among the six locations. The highest mean daily volume (Hennepin Ave. – 2,239) is more than seven times as the lowest mean daily volume (Wirth Parkway – 316) (Table 1), although this ratio is affected by the different number of observations across locations, which include different months of the year and different mixes of weekends and weekdays. Our sample size is large enough for developing models from the pooled counts or counts for each of the six individual locations.

[Table 1 Here]

**Correlates of Traffic Volumes**

Based on the studies described in the literature (e.g., Lindsey et al. (2007), Coutts (2009), Hankey et al. (2012)), we identified 10 independent variables, together with the expected signs, to include in models as correlates of daily trail traffic. These independent variables include neighborhood socio-demographic characteristics, weather and day of week, and characteristics of the built environment of the neighborhood in which the counter is located. Mean values and sources for each variable, together with the dummy variables used are presented in Table 2.

[Table 2 Here]

Two socio-demographic variables, race (*blkpct*) and age (*ynoldpct*), are estimated from the 2010 Census. Two others, education (*collegepct*) and income (*medincthd*), are estimated from the 2000 census due to the lack of the SF3 data in the 2010 Census. The geographic unit for all four socio-demographic variables is the census block group in which the traffic monitor is located (Figure 1). Mixed-mode trail traffic is expected to correlate positively with neighborhood income, education, and proportion of middle-aged population because neighborhoods with these populations may generate more traffic and because trail users from other neighborhoods may prefer to use facilities in neighborhoods with these characteristics.

The built environment variable is population density (*popden*). *Popden* is calculated for the Census block group from the 2010 Census data. Non-motorized traffic volumes are believed to be higher in areas with greater population density because these areas potentially generate more trips.

The four weather variables are retrieved from monthly and daily weather archives in Minneapolis and St. Paul area from the Minnesota Climatology Working Group website. The magnitude of the effect of weather on non-motorized traffic likely varies both by mode and trip purpose. For example, for recreational bicycle and pedestrian trips on trails, the effects of weather are likely to be more similar than the effects on weather on utilitarian bicycle and pedestrian trips on streets and sidewalks, respectively. Mean daily trail traffic is expected to correlate positively with temperature and negatively with precipitation and average wind speed. The direction of the variable *maxdev*, the deviation of daily high temperature from the long-term average of daily high temperatures, is expected to vary by season and direction of variation.

The only temporal variable is a dummy variable *weekend* indicating whether the day of the count is a weekend day or a weekday. Greater volumes of trail traffic are expected on weekend days than on weekdays in any given location because of the time available to individuals to engage in outdoor recreation.

**Model Development and Estimation**

Regression methods used for count data, such as Poisson or negative binomial regression, can be used when the dependent variable is a non-negative integer count rather than a continuous variable. Poisson and negative binomial models naturally fit the characteristics of urban trail traffic better than OLS regression. For instance, negative or fractional traffic cannot occur. In these models, the probability of *y* equals *m* conditioning on the linear combination of $x_1$, $x_2$… and parameter $\lambda$ is given by the following distribution (Long and Freese 2005):

$$P(y = m \mid \lambda, x_1, x_2, ...) = \frac{e^{-\lambda} \lambda^m}{m!} \qquad (2)$$

The choice among types of count models depends on the characteristics of the data. Poisson models are based on the restrictive assumption that the mean equals the variance (i.e., the parameter λ represents both the mean and variance). It has been shown that negative binomial regression is preferred when the variance exceeds the mean (i.e., when over-dispersion exists). In the estimation of negative binomial models, one common assumption is that the mean and the variance of y are $\lambda$ and $\lambda + \alpha\lambda^2$. This approach is called a Type 2 negative binomial model(Cameron and Trivedi 1986). As will be discussed later, over-dispersion exists for our data, indicating that the negative binomial model is preferred to a Poisson model.

Maximum Likelihood Estimation is used to estimate $\alpha$ and the βs of the following model:

$$\ln \lambda = \beta_0 + \beta_1 x_1 + \beta_2 x_2 + ... \qquad (3)$$

The estimated coefficients are labeled as $\hat{\beta}_i$ (i = 1, 2, …). The expected value dependent variable *y* therefore can be predicted as:

$$E(y \mid x_1, x_2, ...) = \hat{\lambda} = \exp(\hat{\beta}_0 + \hat{\beta}_1 x_1 + \hat{\beta}_2 x_2 + ...) \qquad (4)$$

Because the negative binomial function is a nonlinear function, coefficients of models cannot be interpreted as in linear models such as simple OLS. If the coefficient of an independent variable is β, one unit increase of the variable will increase the expected trail count by 100*(exp(β)-1)%, or to their exp(β) times.

Maximum Likelihood Estimation of negative binominal models does not generate the goodness-of-fit statistics, $R^2$, that are generally reported with linear regression. However, scholars have developed statistics known as Pseudo-$R^2$ values that can be used as indexes of the quality of fit. We report McFadden's Pseudo-$R^2$, an index with values ranging from 0 to 1. Higher values of McFadden's Pseudo-$R^2$ indicate a better overall fit for a model but are not interpreted as percentages in the same manner as conventional $R^2$ values (Long and Freese 2005).

We estimate and present here eight different models:

- A general model (Model 1) that incorporates 10 independent socio-demographic, built environment, weather, and temporal variables;
- A six-location model (Model 2) that includes dummy variables for the monitoring locations and omits the socio-demographic and built environment variables; and
- Six trail-specific models (Models 3 – 8), one for each of the six locations where traffic counts have been taken, that include only the weather and temporal variables.

The general model can be used to estimate traffic at locations where traffic counts have not been taken or for proposed new trails because values for the independent variables can be computed for any location on an existing or proposed trail. The six-location model is useful because it includes location dummy variables and can be used to estimate relative traffic volumes among the six locations. The six site-specific models can be used to estimate traffic for each monitoring location when counts are missing or if the TrailMaster © counters are moved to other sites. They can reflect site-specific characteristics of each location which might not be included in the previous two models. We estimated these models using STATA 12© and its extension, SPost 9 (Long and Freese 2005).

**RESULTS**

Models 1 – 8 are presented in Table 3. For each model, the p-value of the over-dispersion test (LR test) is smaller than 0.05. This fact indicates negative binomial regression is preferred.

The fit of the models, as measured by the Pseudo-$R^2$, is comparable, although values are slightly lower for the location specific models, particularly Model 6-Calhoun and Model 5-Cedar. However, the absolute Pseudo- $R^2$ cannot be directly used for judging goodness-of-fit as $R^2$. In our general model (Model 1), the signs on the coefficients of most variables are in the expected direction and all are significant at a less than 1% level. Trail traffic is correlated positively and significantly with neighborhood education, income, proportion of population between 6 and 64, and population density. These results are consistent with those reported by Lindsey et al. (2007) for multiuse trails in Indianapolis, IN.

[Table 3 Here]

Among the weather variables, *tmax* is significant and has the expected positive sign: results indicate that an increase in temperature of one degree Celsius can increase daily trail traffic by 8.5%. The variable *precip* also is significant with the expected sign: an increase of one centimeter of precipitation is associated with a decrease in daily trail traffic by 19.2%. The coefficient for the variable *windavg* is negative, as expected, and significant; a reduction of daily trail traffic by a percentage of 1.7% is expected with an increase in wind speed of 1kph.The variable *maxdev* has a significant negative sign, which indicates one degree more deviation from the 30-year normal temperature will decrease 3.3% of the bicycle traffic in each trail segment. The interpretation of this variable also is as expected: hotter temperatures in summer tend to depress traffic.

The temporal dummy variable, *weekend*, has a significant positive coefficient as expected. On average trail traffic volumes are 34.2% higher on weekends than weekdays.

Our six-location model (Model 2) illustrates the effects of different locations on daily trail traffic counts when controlling for the five weather and temporal variables. The coefficients of all variables also are significant at a less than 1% level. The value of each location dummy variable indicates the daily traffic count relative to that of Wirth Parkway, the location with the lowest mean daily traffic, when the weather and temporal factors are equal. Lake Calhoun Parkway attracts the highest trail traffic volume: 10.8 times the Wirth Parkway volume, followed by Cedar Avenue & Midtown Greenway with 7.6 times, Hennepin Avenue & Midtown Greenway with 6.6 times, Nokomis Parkway with 5.0 times, and West River Parkway & Midtown Greenway with 3.0 times respectively.

Our trail-specific models (Models 3-8) include only the weather variables and the weekend dummy variable. All variables in each of the six models are significant with the expected sign except the weekend variable at the Cedar location on the Midtown Greenway. This unexpected outcome may have to do with unique traffic patterns such as high levels of commuting or utilitarian use relative to recreational use on this section of the Greenway.

**VALIDATION OF MODELS**

To validate our models, we used them to predict trail traffic and compare estimates with actual counts for each monitoring location for one week not included in the dataset, Saturday, September 24, 2011 – Friday, September 30, 2011. For each location, trail traffic was predicted using the general model, the six-location model, and the relevant trail specific model. To illustrate the relative performance of these negative binomial models, we also used OLS regression to predict traffic for each location using the same variables. The predictions for each negative binomial model, together with the actual counts for each location, are graphed in Figure 2. The predicted values generally track the actual values, although, across all models, the divergence between predicted and actual values is greater for the days with higher traffic volumes. The traffic volumes predicted with the general model (Model 1) and the six-location model (Model 2) are quite similar. The performance of the six trail-specific models (Models 3-8) varies across the six monitoring sites, but the trail-specific models do better in tracking the highest traffic volumes at three locations (Calhoun Parkway, Nokomis Parkway and Wirth Parkway).

[Figure 2 Here]

To quantify the magnitude of the error associated with the predictions, we estimated the mean of the absolute value of the difference between the actual traffic volume and the volumes predicted by each model for each site (Table 4). The percentage errors of the grand means for the seven days of validation across the six monitoring locations are similar, ranging from 15.2% for the trail specific models to 17.1% for the six-location model. The percentage error of the grand mean for the general model (16.3%) is between these values.

[Table 4 Here]

The magnitude of error indicated by these grand means masks considerable variation in the mean percentage error for the estimates across models and monitoring sites. Across the six monitoring sites, for example, the mean error for the general model for the validation week ranged from 8.3% to 22.5% (Table 4). The range of error for the validation week across locations for the six-location model was the same. The range of error across locations for the trail-specific models was smaller, from 11.4% to 19.4%, indicating greater consistency in accuracy.

The ranges of mean errors for the validation presented in Table 4 mask the variation in prediction errors for individual days at each monitoring site. For example, on September 27, the traffic count at the Cedar/Midtown Greenway site was 2,252, and the general model estimated daily traffic at the site to be 2,176, a difference of only 3.3%. In comparison, also on September 27, the traffic count on the Nokomis Parkway was 845, but the general model estimated 1,420, an error of 68.1%. For predictions made with the six-location model, the percentage errors for individual days in the validation period across sites range from 3.3% to 68.1%. For the trail-specific models, the comparable range is from essentially zero to 65.3%, indicating predictions from the trail specific models appear to be slightly more consistent.

As is clear in Table 4, the prediction errors associated with OLS models are much greater than the prediction errors resulting from estimation of the general model, the six-location model, and the trail-specific models. The grand means of the OLS percentage errors for general model, the six-location model, and the trail models range from 27.9% to 51.5%, errors that are more than two to three times the errors of the negative binomial models, but consistent with magnitudes presented for similar OLS models previously (Lindsey et al. 2007). The ranges of the OLS prediction errors across monitoring sites and days of the validation week also are much greater.

**APPLICATIONS**

These models have a number of different potential applications, ranging from comparison and validation of correlates of trail traffic reported in other research (e.g., Lindsey 2007) to prediction of trail traffic at street intersections to determine whether crossing improvements are needed to improve safety (e.g., Anderson 2010).These models of mixed-mode trail traffic complement models reported by Hankey et. al (2012) that estimate bicycle and pedestrian traffic levels separately for different types of infrastructure, including streets, bike lanes, and sidewalks. Also, because these trail models are based on more robust datasets from continuous counters, they provide better insight into variations of non-motorized traffic in response to weather than models estimated from two-hour counts taken only seasonally (e.g., Hankey et al. 2012).

For practical applications, these new trail traffic models may be applied in different circumstances, depending on the needs of planners and managers. The general model, which incorporates variables to describe neighborhood characteristics, can be applied to any point on an existing trail or to a location where a trail might be developed, assuming data from the Census and other sources can be obtained for the location of interest. The six-location model, which replaces neighborhood variables with dummy variables for locations, can be used to estimate relative traffic for each monitoring locations. It will be helpful especially when we compare different locations. The trail specific models may be useful for imputing missing observations from the infrared counters, so planners can obtain yearly totals and estimate total miles traveled on trails. An advantage of the trail specific models is that they can control for factors not included in the previous two models.

From a pragmatic perspective, the usefulness of these models depends on the application and the tolerance for uncertainty. For example, if the Minneapolis Department of Public Works needed to identify trail-street intersections with highest traffic volumes, the general model could be easily applied for each such intersection in the city to provide a list for further field screening. Similarly, if the Park and Recreation Board were considering investment in several new trails, and if potential use were a decision-criterion, then the general model could be helpful in ranking alternatives. In these types of applications, use of a consistent approach helps to ameliorate the significance of error associated with the estimates. Other potential applications of these models include allocation of maintenance dollars, identification of the need for new signage or other safety treatments, and targeting of public safety campaigns. A mean error of 16% generally will be acceptable when order-of-magnitude estimates are sufficient for decision-making or when no data at all are available, which historically has been the case for many proposed trail projects. To place this modeling error in perspective, motorized vehicle pneumatic tube counters are considered to be working properly if automated counts are within +10% of actual (Turner et al. 2010), while error rates for automated inductive loop counters for bicycles may be lower or higher depending on location, method of installation, and degree of maintenance (Nordback and Janson 2010; Nordback et al. 2011).

**CONCLUSIONS**

Use of negative binomial regression models is appropriate when dependent variables are non-negative counts with over-dispersion. These models can be used to estimate non-motorized – bicycle and pedestrian – traffic on urban multiuse trails. Using traffic counts from active infrared-monitors at six locations on trails in Minneapolis, we estimated a general model with 10 variables that have been reported to influence trail traffic, a six-location model that includes dummy variables for each monitoring site rather than neighborhood and built environment variables, and six trail-specific models that include only weather variables and a dummy variable for a weekend day as controls. All the models have reasonable fit, and all the variables in the models are significant. Our models demonstrate significant correlation of social-economic and built environment characteristics with multimodal trail traffic. They also confirm the significant and consistent effects of weather and day of week on trail traffic counts.

To validate models we predicted trail traffic for each monitoring site for a week not included in the data set from which the models originally were estimated. As measured by the grand mean of prediction errors across locations and days of the validation week, the prediction errors for the general model, the six-location model, and the trail-specific models were comparable, approximately 15-17%. The ranges of prediction errors across days and locations for the trail-specific models were smaller, indicating they may be more consistent. The models clearly outperform the same models estimated with OLS regression. In general, the mean percentage errors of predictions with negative binomial models are more consistent and only 1/2-1/3 of the standard OLS models.

These models can be used in many different planning and management scenarios. A limitation of this research arises from the data set used in the analysis. Data are available for only six locations and, across these locations, for unequal time periods. A larger dataset that includes data throughout the calendar year for all sites would make the analyses and models more robust by incorporating extremely high and low counts. Increasing the number of monitor locations potentially could make the data more representative of other trail sites in Minneapolis or throughout the metropolitan region. The fact that the active infrared counts do not distinguish bicycles and pedestrians limits some applications of the models model: they will be most useful in decision making contexts where total traffic and not mode-specific traffic is most relevant. As new technologies that provide separate counts of bicyclists and pedestrians are more widely deployed, mode-specific data will become available, and mode-specific models can be developed.

**ACKNOWLEDGEMENTS**

We thank Shaun Murphy and Simon Blensky of the Minneapolis Department of Public Works and Jennifer Ringold and Ginger Cannon of the Minneapolis Park and Recreational

Board for their help and facilitation in the collection of original data. We also thank the anonymous referees for their valuable comments.

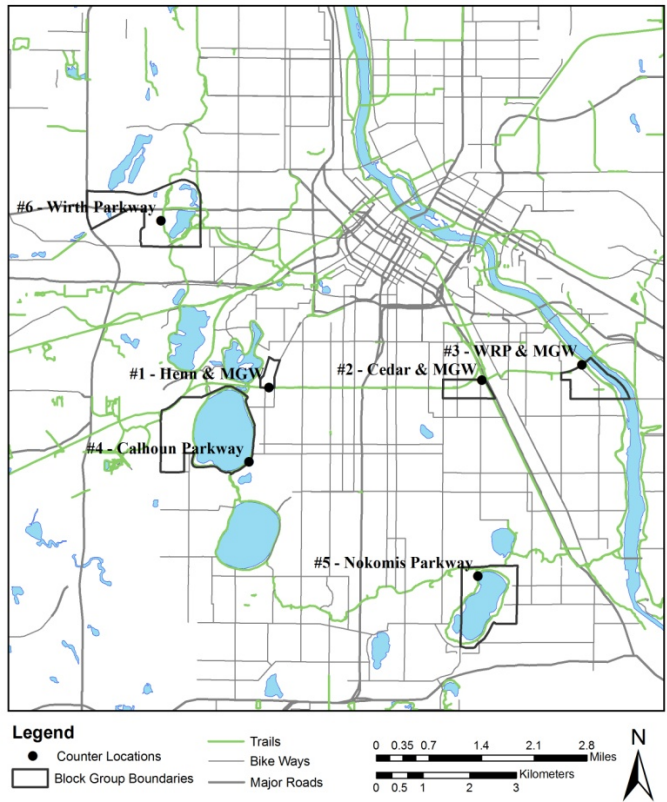

1
2         **Fig. 1.** Minneapolis Urban Trail Counting Locations
3
4

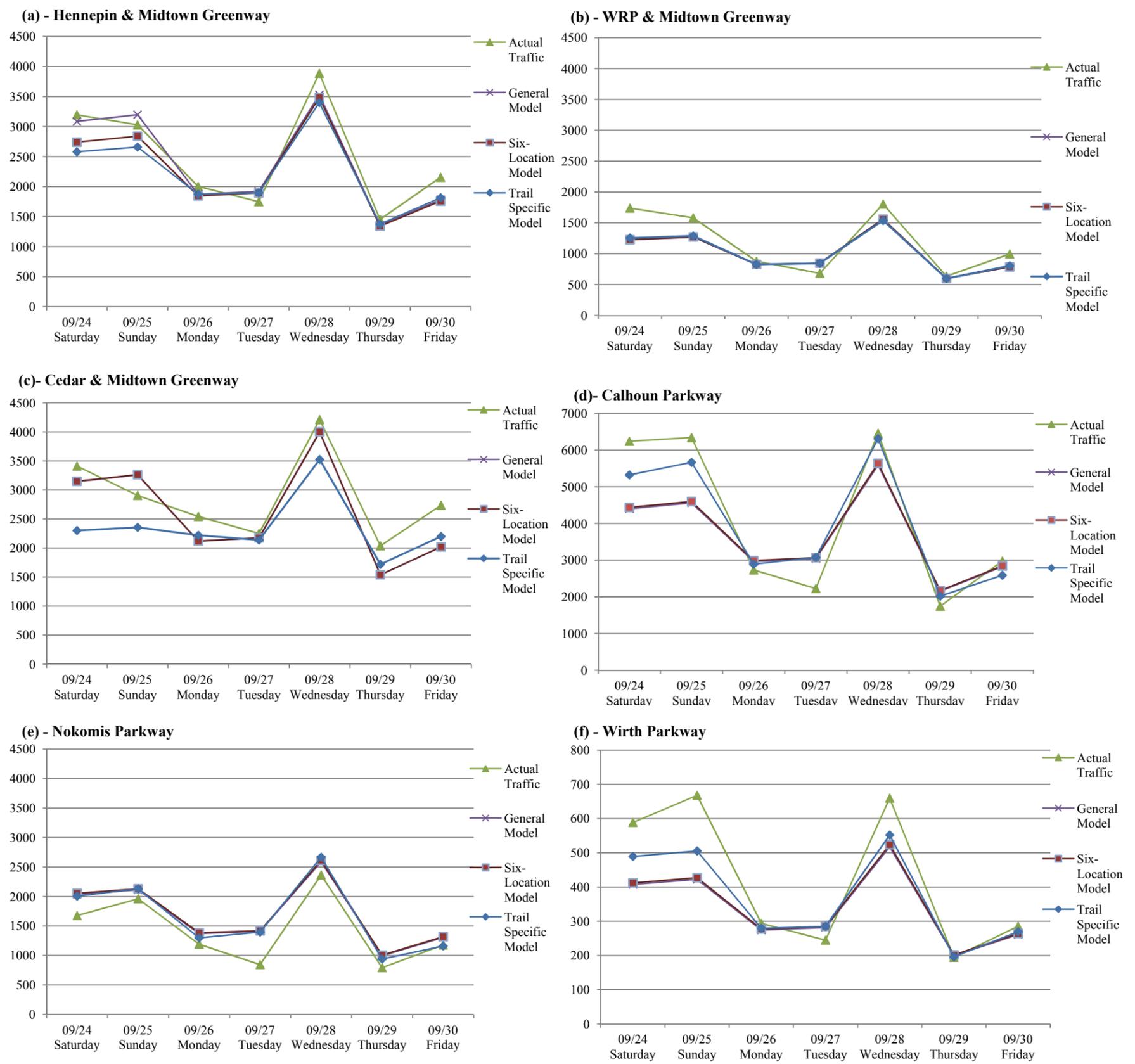

**Fig.2.** Predicted and Actual Trail Traffic, September 24-30, 2011

**Table 1.** Mean daily non-motorized traffic volumes on Minneapolis multiuse trails.

| Location | Mean Daily | Days of Counts | Starting Date | Ending Date |
|---|---|---|---|---|
| (1) Hennepin Ave. & Midtown Greenway (MGW) | 2,239 | 427 | 06/21/2010 | 09/23/2011 |
| (2) West River Pkwy & MGW | 1,031 | 405 | 07/07/2010 | 09/23/2011 |
| (3) Cedar Ave. & MGW | 2,010 | 272 | 11/10/2010 | 09/23/2011 |
| (4) Lake Calhoun Parkway | 3,370 | 269 | 12/10/2010 | 09/23/2011 |
| (5) Lake Nokomis Parkway | 1,452 | 261 | 12/10/2010 | 09/23/2011 |
| (6) Wirth Parkway | 316 | 264 | 12/11/2010 | 09/23/2011 |

**Table 2.** Variables Selected for Model 1 Building and Expected Signs

| Variables | Notes | Mean | Expected Sign |
|---|---|---|---|
| *Neighborhood Socio-demographic Characteristics* | | | |
| *blkpct* | Percentage of African American residents. | 6.099 | - |
| *collegepct* | Percentage of residents with college education* | 59.79 | + |
| *yngoldpct* | Percentage of population over 64 or below 6 | 18.67 | - |
| *medincthd* | Median household income. (1,000 dollars) | 46.01 | + |
| *Neighborhood Built Environment* | | | |
| *popden* | Population density (per square kilometer). | 2288 | + |
| *Weather Conditions*** | | | |
| *tmax* | Recorded high temperature. (in Celsius) | 14.06 | + |
| *maxdev* | Deviation from the 30-year normal temperature. | 0.533 | +/- |
| *precip* | Precipitation. (centimeters) | 0.254 | - |
| *windavg* | Average wind speed. (kph) | 13.69 | - |
| *Temporal Dummies* | | | |
| *weekend* | Saturday or Sunday (equals 1, otherwise 0) | 0.288 | + |
| *Location Dummies* | | | |
| *henn* | Hennepin @ MGW counter | 0.225 | +/- |
| *wrp* | West River Parkway @ MGW counter | 0.213 | +/- |
| *cedar* | Cedar @ MGW counter | 0.143 | +/- |
| *calhoun* | Lake Calhoun counter | 0.142 | +/- |
| *nokomis* | Lake Nokomis counter | 0.138 | +/- |

\* Educational attainment data are only available for the people over 25 years old.
\*\* from Minnesota Climatology Working Group
(http://www.weather.umn.edu/doc/prelim_lcd_msp.htm)

Table 3. Estimation Results of the Models

| Variables | (1) General Model n=1898 | (2) Six-location Model n=1898 | Trail-specific Models 3-8 | | | | | |
|---|---|---|---|---|---|---|---|---|
| | | | (3) Hennepin n=427 | (4) WRP n=405 | (5) Cedar n=272 | (6) Calhoun n=269 | (7) Nokomis n=261 | (8) Wirth n=264 |
| **Pseudo-R²** | 0.1329 | 0.1329 | 0.1162 | 0.1283 | 0.1111 | 0.0986 | 0.1197 | 0.1596 |
| (Constant) | -150.5*** | 4.331*** | 6.221*** | 5.397*** | 6.448*** | 6.611*** | 6.029*** | 4.166*** |
| *Social Demographic Characteristics* | | | | | | | | |
| blkpct | **4.132*** | - | - | - | - | - | - | - |
| collegepct | **0.701*** | - | - | - | - | - | - | - |
| yngoldpct | **-0.195*** | - | - | - | - | - | - | - |
| medincthd | **1.650*** | - | - | - | - | - | - | - |
| *Built Environment* | | | | | | | | |
| popden | **0.007*** | - | - | - | - | - | - | - |
| *Climate Conditions* | | | | | | | | |
| tmax | **0.082*** | **0.082*** | **0.083*** | **0.085*** | **0.077*** | **0.087*** | **0.074*** | **0.093*** |
| maxdev | **-0.033*** | **-0.033*** | **-0.039*** | **-0.040*** | **-0.044*** | **-0.017*** | -0.008 | **-0.043*** |
| precip | **-0.214*** | **-0.214*** | **-0.190*** | **-0.221*** | **-0.218*** | **-0.235*** | **-0.216*** | **-0.224*** |
| windavg | **-0.016*** | **-0.017*** | **-0.015*** | **-0.017*** | **-0.011*** | **-0.020*** | **-0.019*** | **-0.018*** |
| *Temporal Dummy* | | | | | | | | |
| weekend | **0.294*** | **0.294*** | **0.202*** | **0.282*** | -0.076 | **0.571*** | **0.417*** | **0.423*** |
| *Location Dummies* | | | | | | | | |
| henn | - | **1.894*** | - | - | - | - | - | - |
| wrp | - | **1.091*** | - | - | - | - | - | - |
| cedar | - | **2.033*** | - | - | - | - | - | - |
| calhoun | - | **2.377*** | - | - | - | - | - | - |
| nokomis | - | **1.607*** | - | - | - | - | - | - |
| *Dispersion Factor* | | | | | | | | |
| *p in LR test* | 0.000 | 0.000 | 0.000 | 0.000 | 0.000 | 0.000 | 0.000 | 0.000 |

Note: The coefficients with bold numbers are consistent with expected signs. Coefficients which are significant at 0.1 level, 0.05 level and 0.01 level are labeled as *, **, and ***.

Table 4. Predicted and Actual Trail Traffic, September 24-30, 2011

| Site | Model Type | Mean Daily Traffic | Model 1 General | | Model 2 Six-Location | | Model 3-8 Trail Specific | |
|---|---|---|---|---|---|---|---|---|
| | | | Predict | Error | Predict | Error | Predict | Error |
| Hennepin | NB2 | 2496 | 2393 | 8.3 | 2271 | 10.5 | 2229 | 11.4 |
| | OLS | | 2703 | 19.4 | 2670 | 18.3 | 2760 | 19.4 |
| WRP | NB2 | 1188 | 1014 | 17.2 | 1017 | 17.0 | 1022 | 16.5 |
| | OLS | | 1454 | 27.3 | 1458 | 27.7 | 1277 | 20.6 |
| Cedar | NB2 | 2871 | 2606 | 13.8 | 2610 | 13.7 | 2351 | 17.3 |
| | OLS | | 2730 | 10.1 | 2732 | 10.2 | 2843 | 9.9 |
| Calhoun | NB2 | 4103 | 3649 | 20.7 | 3679 | 20.7 | 3982 | 14.3 |
| | OLS | | 4033 | 44.1 | 4037 | 44.2 | 4704 | 38.0 |
| Nokomis | NB2 | 1430 | 1689 | 22.5 | 1703 | 23.5 | 1657 | 19.4 |
| | OLS | | 2082 | 55.9 | 2085 | 56.2 | 1975 | 47.1 |
| Wirth | NB2 | 419 | 338 | 17.4 | 342 | 17.1 | 368 | 12.1 |
| | OLS | | 1048 | 151.5 | 1051 | 152.6 | 471 | 32.6 |
| **Grand Mean Error NB (%)** | | | 16.6 | | 17.1 | | 15.2 | |
| **Grand Mean Error OLS (%)** | | | 51.4 | | 51.5 | | 27.9 | |

Note: All errors are percentages in absolute value.